\documentclass[aps]{revtex4}
\usepackage{amssymb,epsf}

\begin{document}

\title{Nonsingular Universes in Gauss-Bonnet Gravity's Rainbow}
\author{Seyed Hossein Hendi$^{1,2}$\footnote{
email address: hendi@shirazu.ac.ir}, Mehrab Momennia$^{1}$\footnote{%
email address: m.momennia@shirazu.ac.ir}, Behzad Eslam Panah$^{1}$\footnote{%
email address: behzad.eslampanah@gmail.com} Mir Faizal$^{3}$\footnote{%
email address: f2mir@uwaterloo.ca}}
\affiliation{$^{1}$ Physics Department and Biruni Observatory, College of Sciences,
Shiraz University, Shiraz 71454, Iran\\
$^2$ Research Institute for Astronomy and Astrophysics of Maragha (RIAAM),
Maragha, Iran\\
$^3$ Department of Physics and Astronomy, University of Lethbridge,
Lethbridge, AB T1K 3M4, Canada}

\begin{abstract}
In this paper, we will study the rainbow deformation of the FRW cosmology in
both Einstein gravity and Gauss-Bonnet gravity. We will demonstrate that the
singularity in the FRW cosmology can be removed because of the rainbow
deformation of the FRW metric. We will obtain the general constraints
required for the FRW cosmology to be free from singularities. It will be
observed that the inclusion of Gauss-Bonnet gravity can significantly change
the constraints required to obtain a nonsingular universes. We will use a
rainbow functions motivated from the hard spectra of gamma-ray bursts to
deform the FRW cosmology, and it will be explicitly demonstrated that such a
deformation removes the singularity in the FRW cosmology.
\end{abstract}

\maketitle

\section{Introduction}

Even though it has not been possible to construct a quantum theory of
gravity, there are many proposals for quantum gravity, and such proposals
can have interesting physical consequences \cite%
{Amelinob,Amelinoc,MogueijoSb,Smolin}. In fact, many of these proposals have
predict similar physical consequences, and one such almost universal
prediction of many different approach to quantum gravity is the deformation
of the standard relativistic dispersion relation \cite{Kempf,Brau}. Such a
deformation of the standard relativistic dispersion relation occurs in
various different approaches to quantum gravity, such as the spacetime
discreteness \cite{Hooft}, spacetime foam models \cite{AmelinoEMNS},
spontaneous symmetry breaking of Lorentz invariance in string field theory
\cite{Kostelecky}, and spin-network in Loop quantum gravity \cite{Cambini}.
The standard energy-momentum dispersion relation gets deformed to a modified
dispersion relation (MDR) near the Planck scale. It is possible to use MDR
to explain certain astronomical and cosmological observations, such as the
threshold anomalies of ultra high energy cosmic rays and TeV photons \cite%
{AmelinoLNa,Colladay,Takeda,Finkbeiner,Myers,Jacobson,Amelinoe}.

The MDR is based on the existence of a maximum energy scale, and it has been
possible to construct a theory with such an intrinsic maximum energy scale
\cite{Amelinoc,MogueijoSb}. This theory is called the doubly special
relativity, and in this theory the Planck energy ($E_{P}$) and the velocity
of light ($c$) are two universally invariant quantities. Just as it is not
possible for a particle to attain a velocity greater than the velocity of
light in special relativity, it is not possible for a particle to attain an
energy larger than the Planck energy in doubly special relativity. In the
doubly special relativity, the Lorentz transformations are deformed to a set
of nonlinear Lorentz transformation in momentum space. In fact, this
deformation of the Lorentz transformations directly deforms the standard
energy-momentum relation. It has been possible to extend the doubly special
relativity to curved spacetime and obtain doubly general relativity \cite%
{MogueijoSc}. In this theory, one assumes that the geometry of spacetime
depends on the energy of the test particle. So, we do not have a single
metric describing the geometry of spacetime, but a one-parameter family of
energy dependent metrics. These metrics depend on the energy of the test
particles. As we have a family of energy dependent metrics in such a theory,
this theory is called the gravity's rainbow \cite{MogueijoSc, AmelinoEMNS}.

Recently, gravity's rainbow has been used to study the high energy behavior
of various physical systems \cite%
{GalanM,Hackett,Aloisio,LingLZ,GarattiniM,Chang,Santos,AliK}. The rainbow
deformation of various black hole solutions have been performed, and their
properties have been studied \cite%
{GalanMb,Ali,GimKa,GimKb,MuWY,HendiFEP,HendiPEM}. The hydrostatic
equilibrium for compact objects and the structure of neuron stars has also
been investigated using the gravity's rainbow \cite{HendiBEP,GarattiniMG}.
Furthermore, the effects of gravity's rainbow on wormholes have also been
investigated \cite{GarattiniL}. The gravity's rainbow has also been used for
analyzing the effects of rainbow functions on gravitational force and
Starobinsky model of $f(R)$ gravity \cite{Sefiedgar,Chatrabhuti}.

It may be noted that string theory can be regarded as two dimensional
theory, and the target space metric can be regarded as a matrix of coupling
constants for this two dimensional theory. These coupling constants will
flow due to the renormalization group flow, and so the target space metric
will depend on the scale at which spacetime is probed, but this scale would
in turn depend on the energy of the test particle used to probe this
spacetime. Thus, the target space metric in string theory would depend on
the energy of the probe, and so gravity's rainbow is motivated from string
theory \cite{MogueijoSc}. It may be noted that the low energy effective
field theory approximation to the heterotic string theory \cite%
{Gross,MetsaevTa}, produces the Gauss-Bonnet (GB) gravity \cite{Stelle,Maluf}%
. It has been demonstrated that the low-energy expansion of string theory
effective action contains the GB term and a scalar field \cite{MetsaevTb}.
It is possible to neglect the effect of this scalar field as it can be
regarded as a constant field. The GB gravity contains curvature-squared
terms, and is free of ghosts. Furthermore, the corresponding field equations
contain no more than second derivatives of the metric \cite%
{BoulwareD,Zumino,Callan,Cai}. Black object solutions have also been studied
in GB gravity \cite%
{MyersS,ChoN,Mignemi,ChenGO,Bogdanos,HEndiE,CaiLS,Brihaye,Gaete,Ayzenberg}.
As both GB gravity and gravity's rainbow can be motivated from string
theory, there is a strong motivation to study rainbow deformation of GB
gravity. In fact, the thermodynamics of black holes has been studied using a
combination of the gravity's rainbow and GB gravity \cite{HendiF}. We also
note that GB gravity has been used for analyzing various cosmological models
\cite%
{Deruelle,KimM,Amendola,Leith,Elizalde,BrihayeR,ChingangbamSTT,Bamba,Capozziello,Kanti}%
. The FRW cosmology in Einstein gravity's rainbow has also been analyzed
\cite{AwadAM,Ling}. It was observed that the universe is nonsingular in this
model, however, it is important to analyze other cosmological models, so
that we can know if this is a model dependent effect or a general feature of
gravity's rainbow. Furthermore, no work has been done on cosmological
applications of GB-gravity's rainbow, even though there are strong string
theoretical motivations to perform this analysis. So, in this paper, we are
going to analyze a cosmological model using GB-gravity's rainbow. It will be
observed that this model is also nonsingular, and so it seems that removal
of big bang singularity because of the rainbow deformation is a general
feature of rainbow deformation of any model of gravity.

\section{FRW RAINBOW COSMOLOGY IN EINSTEIN GRAVITY \label{FRWEINSTEIN}}

Here, we are going to modify FRW universe in the Einstein gravity's rainbow.
We consider the Lagrangian of Einstein gravity with a matter field as
\begin{equation}
\mathcal{L}_{E}=\mathcal{R}+\mathcal{L}_{m},  \label{LE}
\end{equation}%
where $\mathcal{R}$ is the Ricci scalar and $\mathcal{L}_{m}$ is the
Lagrangian of matter. Variation of the action (\ref{LE}) with respect to the
metric tensor $g_{\mu \nu }$ leads to
\begin{equation}
G_{\mu \nu }^{E}=8\pi GT_{\mu \nu },  \label{FieldeqE}
\end{equation}%
where $G_{\mu \nu }^{E}=R_{\mu \nu }-\frac{1}{2}Rg_{\mu \nu }$\ is the
Einstein tensor. The energy-momentum tensor can be expressed as
\begin{equation}
T_{\mu \nu }=\rho u_{\mu }u_{\nu }+P(g_{\mu \nu }+u_{\mu }u_{\nu }),
\label{EMTensor}
\end{equation}%
where $\rho $ and $P$ are the energy density and the pressure of perfect
fluid, respectively. Here $u_{\mu }$ is defined as
\begin{equation}
u_{\mu }=(f^{-1}(\varepsilon),0,0,0,0),
\end{equation}%
and it is a unit vector
\begin{equation}
g^{\mu \nu } u_{\mu }u_{\nu }=-1.
\end{equation}

Now, as we want to analyze this model using gravity's rainbow, we will first
review gravity's rainbow. The gravity's rainbow is based on the deformation
of the standard energy-momentum dispersion relation,
\begin{equation}
E^{2}f(\varepsilon)^{2}-p^{2}g(\varepsilon)^{2}=m^{2},  \label{DeforEM}
\end{equation}%
where $\varepsilon=E/E_{p}$ and the functions $f(\varepsilon)$ and $%
g(\varepsilon)$ are called rainbow functions, and $m$ is the mass of the
test particle. In the IR limit, we have $lim_{\varepsilon\rightarrow
0}f(\varepsilon)=lim_{\varepsilon\rightarrow 0}g(\varepsilon)=1$ and so the
standard energy-momentum dispersion relation is recovered in the IR limit of
this theory. Thus, the gravity's rainbow reduces to standard general
relativity in the IR limit. As the Planck energy is the largest energy that
a particle can attain, we can write
\begin{equation}
\varepsilon\leq 1.
\end{equation}

The exact forms of the rainbow functions are constructed using various
theoretical and observational motivations. In fact, the study that was done
on the hard spectra from gamma-ray bursts has been used as a motivation to
construct the following rainbow functions \cite{AmelinoEMNS}
\begin{equation}
f(\varepsilon)=\frac{e^{\varepsilon}-1}{\varepsilon}\ \ \ \ \ \ \&\ \ \ \ \
g(\varepsilon)=1.  \label{MDR}
\end{equation}
Now after substituting it in Eq. (\ref{DeforEM}), we can find the
corresponding MDR
\begin{equation}
p^{2}=E^{2}\left( \frac{e^{\varepsilon}-1}{\varepsilon}\right) ^{2}.
\label{MDR1}
\end{equation}

In order to compare the results of Einstein gravity with GB theory, one
should reformulate them with identical dimensions. Since the GB term does
not contribute in four dimensions, we consider the following five
dimensional spacetime
\begin{equation}
ds^{2}=-\frac{dt^{2}}{f(\varepsilon )^{2}}+\frac{R(t)^{2}}{g(\varepsilon
)^{2}}dx_{i}^{2},~~\ \ \ \ \ \ i=1,2,3,4,  \label{metric}
\end{equation}%
where $R(t)$ is scale factor, and we consider a flat universe with $k=0$.
Using the above metric, field equation (\ref{FieldeqE}) and the
energy-momentum tensor\ (\ref{EMTensor}), the FRW equations in Einstein
gravity's rainbow can be written as
\begin{equation}
\left( H-\frac{\dot{g}(\varepsilon )}{g(\varepsilon )}\right) ^{2}+\frac{%
\dot{g}(\varepsilon )}{g(\varepsilon )}\left( H-\frac{3\dot{g}(\varepsilon )%
}{4g(\varepsilon )}\right) =\frac{4\pi G\rho }{3f(\varepsilon )^{2}},
\label{FEQE1}
\end{equation}%
\begin{eqnarray}
&&g(\varepsilon )\left[ \dot{H}-\frac{\ddot{g}(\varepsilon )}{2g(\varepsilon
)}-\frac{\dot{g}(\varepsilon )}{g(\varepsilon )}\left( H-\frac{\dot{g}%
(\varepsilon )}{g(\varepsilon )}\right) +\left( H+\frac{\dot{f}(\varepsilon )%
}{2f(\varepsilon )}\right) \left( 2H-\frac{\dot{g}(\varepsilon )}{%
g(\varepsilon )}\right) \right]  \nonumber \\
&& \\
&&-2\left( H-\frac{\dot{g}(\varepsilon )}{g(\varepsilon )}\right) ^{2}+\frac{%
\dot{g}(\varepsilon )}{g(\varepsilon )}\left( \frac{3\dot{g}(\varepsilon )}{%
2g(\varepsilon )}-2H\right) =-\frac{8\pi G(\rho +P)}{3f(\varepsilon )^{2}},
\label{FEQE2}
\end{eqnarray}%
which $H=\dot{R(t)}/R(t)$ is the Hubble parameter. It may be noted that we
have used the notations $\dot{A}=\frac{dA}{dt}$ and $\ddot{A}=\frac{d^{2}A}{%
dt^{2}}$. The conservation of energy-momentum tensor can be written as
\[
\nabla _{\mu }T_{\nu }^{\mu }=\partial _{\mu }T_{\nu }^{\mu }-\Gamma _{\mu
\nu }^{\lambda }T_{\lambda }^{\mu }+\Gamma _{\mu \lambda }^{\mu }T_{\nu
}^{\lambda }=0,
\]%
and this equation reduces to
\begin{equation}
\dot{\rho}+2\left( 2H-\frac{\dot{g}(\varepsilon )}{g(\varepsilon )}\right)
(\rho +P)=0.  \label{ConsEqE}
\end{equation}

We can consider a large range of ultra relativistic particles, which are in
thermal equilibrium with an average energy $\epsilon \sim T$. The continuity
equation leads to the first law of thermodynamics (as in standard cosmology)
\begin{equation}
d(\rho V)=-PdV,  \label{PV}
\end{equation}%
where $V$ is the volume and $V=[R(t)/g(\varepsilon)]^{4}$. The Eq. (\ref{PV}%
) along with the integrability condition $\frac{\partial ^{2}S }{\partial
V\partial P}=\frac{\partial ^{2}S}{\partial P\partial V}$ \cite{Kolb} leads
to a constant entropy
\begin{equation}
S=\frac{V(\rho +P)}{T}=const.
\end{equation}

In this paper, we are going to consider the following equation of state
(EoS)
\begin{equation}
P=(\gamma -1)\rho.  \label{EoS}
\end{equation}
The FRW spacetime is singular at $t=0$, if this EoS is used in the standard
cosmology. In the above equation, $\gamma$ is the EoS parameter. For this
pressure, the average energy $\epsilon$ can be written as
\begin{equation}
\epsilon \sim T=c\gamma \rho V,
\end{equation}%
where $T$ is temperature and $c$ is a constant (which is equal to $1/S$).
Using the EoS (\ref{EoS}) in Eq. (\ref{ConsEqE}), we obtain the following
equation
\begin{equation}
\frac{d\rho }{d\ln [R(t)^{2}/g(\varepsilon)]}=-2\gamma \rho ,
\end{equation}%
which can be solved to give a density $\rho=[R(t)^{2}/g(\varepsilon)]^{-2%
\gamma}$. This leads to an average energy
\begin{equation}
\epsilon =\frac{c \gamma }{R(t)^{4}}\rho ^{\frac{\gamma -2}{\gamma }}.
\label{EnergyE}
\end{equation}

\section{MODIFIED FRW RAINBOW COSMOLOGY IN GB GRAVITY \label{FRWGB}}

Now, we are going to analyze the FRW universe using gravity's rainbow with
GB term. We are also going to study its effect on the early universe using a
semi-classical approximation. The Lagrangian of Einstein-GB gravity can be
written as
\begin{equation}
\mathcal{L}_{tot}=\mathcal{R}+\alpha \mathcal{L}_{GB}+\mathcal{L}_{m},
\label{Ltot}
\end{equation}%
where the parameter $\alpha$ in the second term of Eq. (\ref{Ltot}) is the
GB coefficient with dimension $(length)^{2}$, and $\mathcal{L}_{GB}$ is the
Lagrangian of GB gravity
\begin{equation}
\mathcal{L}_{GB}=R_{\mu \nu \tau \sigma }R^{\mu \nu \tau \sigma }-4R_{\mu
\nu }R^{\mu \nu }+\mathcal{R}^{2}.
\end{equation}

Variation of the action (\ref{Ltot}) with respect to the metric $g_{\mu \nu
} $ leads to
\begin{equation}
G_{\mu \nu }^{E}+\alpha G_{\mu \nu }^{GB}=8\pi GT_{\mu \nu },
\label{Fieldeq}
\end{equation}%
where $G_{\mu \nu }^{GB}=2(R_{\mu \tau \sigma \lambda }R_{\nu }^{\tau \sigma
\lambda }-2R_{\mu \tau \nu \sigma }R^{\tau \sigma }-2R_{\mu \lambda }R_{\nu
}^{\lambda }+\mathcal{R}R_{\mu \nu })-\frac{1}{2}\mathcal{L}_{GB}g_{\mu \nu
} $. Using the metric (\ref{metric}), field equation (\ref{Fieldeq}), and
the energy-momentum tensor (\ref{EMTensor}), we can write the FRW equations
in gravity's rainbow with GB term as
\begin{equation}
\left( H-\frac{\dot{g}(\varepsilon )}{g(\varepsilon )}\right) ^{2}+\frac{%
\dot{g}(\varepsilon )}{g(\varepsilon )}\left( H-\frac{3\dot{g}(\varepsilon )%
}{4g(\varepsilon )}\right) +\alpha \frac{f(\varepsilon )^{2}}{8}\left( 2H-%
\frac{\dot{g}(\varepsilon )}{g(\varepsilon )}\right) ^{4}=\frac{4\pi G\rho }{%
3f(\varepsilon )^{2}},  \label{FEQGB1}
\end{equation}%
\begin{eqnarray}
&&g(\varepsilon )\left[ \dot{H}-\frac{\ddot{g}(\varepsilon )}{2g(\varepsilon
)}-\frac{\dot{g}(\varepsilon )}{g(\varepsilon )}\left( H-\frac{\dot{g}%
(\varepsilon )}{g(\varepsilon )}\right) +\right. \left. \left( H+\frac{\dot{f%
}(\varepsilon )}{2f(\varepsilon )}\right) \left( 2H-\frac{\dot{g}%
(\varepsilon )}{g(\varepsilon )}\right) \right] +\frac{\dot{g}(\varepsilon )%
}{g(\varepsilon )}\left( \frac{3\dot{g}(\varepsilon )}{2g(\varepsilon )}%
-2H\right)  \nonumber \\
&&  \nonumber \\
&&-2\left( H-\frac{\dot{g}(\varepsilon )}{g(\varepsilon )}\right) ^{2}+\frac{%
\alpha }{2}g(\varepsilon )f(\varepsilon )^{2}\left( 2H-\frac{\dot{g}%
(\varepsilon )}{g(\varepsilon )}\right) ^{2}\left[ 2(\dot{H}+H^{2})+\frac{%
\dot{g}(\varepsilon )^{2}}{2g(\varepsilon )^{2}}+\left( 1-\frac{1}{%
2g(\varepsilon )}\right) \left( 2H-\frac{\dot{g}(\varepsilon )}{%
g(\varepsilon )}\right) ^{2}\right.  \nonumber \\
&&  \nonumber \\
&&\left. -\frac{\ddot{g}(\varepsilon )}{g(\varepsilon )}-\left( 2H-\frac{%
\dot{f}(\varepsilon )}{f(\varepsilon )}\right) \left( 2H-\frac{\dot{g}%
(\varepsilon )}{g(\varepsilon )}\right) \right] =-\frac{8\pi G(\rho +P)}{%
3f(\varepsilon )^{2}}.  \label{FEQGB2}
\end{eqnarray}%
Here for $\alpha =0$, Eqs. (\ref{FEQGB1}) and (\ref{FEQGB2}) reduce to Eqs. (%
\ref{FEQE1})\ and (\ref{FEQE2}), respectively. The conservation equation for
GB-gravity's rainbow can be written as
\begin{equation}
\dot{\rho}+2\left( 2H-\frac{\dot{g}(\varepsilon )}{g(\varepsilon )}\right)
(\rho +P)=0.  \label{ConsEqGB}
\end{equation}

It may be noted that Eq. (\ref{ConsEqGB}) is the same as conservation
equation obtained in Einstein gravity's rainbow, Eq. (\ref{ConsEqE}). Now,
using the same procedure with Eqs. (\ref{EoS}) and (\ref{ConsEqGB}), it can
be demonstrated that the average energy has the same form for GB gravity (%
\ref{EnergyE}),
\begin{equation}
\epsilon =\frac{c \gamma }{R(t)^{4}}\rho ^{\frac{\gamma -2}{\gamma }}.
\label{EnergyGB}
\end{equation}

\section{WHEN A NONSINGULAR RAINBOW UNIVERSE IN GB GRAVITY IS POSSIBLE?
\label{POSSIBILITY}}

Before analyzing how the MDR (Eq. (\ref{MDR1})) leads to a nonsingular
cosmology, it is useful to discuss the general conditions on the rainbow
functions that lead to a nonsingular universe. Substituting Eq. (\ref{EoS})
and the modified Friedmann equation of gravity's rainbow (Eq. (\ref{FEQGB1}%
)) in the conservation equation (Eq. (\ref{ConsEqGB})), we obtain
\begin{equation}
\dot{\rho}=\pm 2\gamma \rho \left\{ \frac{8}{\alpha f(\varepsilon )^{2}}%
\left[ \frac{4\pi G\rho }{3f(\varepsilon )^{2}}-H\frac{\dot{g}(\varepsilon )%
}{g(\varepsilon )}-\left( H-\frac{\dot{g}(\varepsilon )}{g(\varepsilon )}%
\right) ^{2}\right] \right\} ^{\frac{1}{4}}.  \label{ConsVSfg}
\end{equation}

A similar system has been studied in Ref. \cite{Awad}, and this analysis was
performed using the Hubble rate. However, it is also possible to study this
model using density $\rho$ instead of the Hubble rate $H$ \cite{AwadAM}. Our
analysis will be based on the approach used in Ref. \cite{AwadAM}, and we
will demonstrate that this cosmological model is free from finite-time
singularities. This is because an upper bound for the density $\rho$ is
reached in an infinite time. Thus, there is a point at which the density
diverges, however, that point exists at an infinite time.

To use this explanation, we need a differential equation for $\rho $ with
respect to time. If we write $f(\varepsilon)$ and $g(\varepsilon)$ as
functions of $\rho$ and $R$ instead of $\epsilon$ according to Eq. (\ref%
{EnergyGB}), the equation (\ref{ConsVSfg}) will be too complicated, and we
will not be able to solve it. So, to solve this problem, we choose the
following form of the solution (separation of variables)
\begin{equation}
g(\varepsilon,t)=G(\varepsilon)R(t),  \label{Choice}
\end{equation}%
and it has the following consequence
\begin{equation}
\frac{\dot{g}(\varepsilon)}{g(\varepsilon)}=\frac{\dot{g}(\varepsilon,t)}{%
g(\varepsilon,t)}=\frac{G(\varepsilon)\dot{R(t)}}{G(\varepsilon)R(t)}=\frac{%
\dot{R(t)}}{R(t)}=H.
\end{equation}
According to Eq. (\ref{Choice}), the first deformed FRW equation (\ref%
{FEQGB1}) reduces to
\begin{equation}
H^{2}+\frac{1}{2}f(\varepsilon)^{2}H^{4}\alpha =\frac{16\pi G\rho }{%
3f(\varepsilon)^{2}}.  \label{FEQGBChoice}
\end{equation}
In addition, considering Eq. (\ref{Choice}), one can show that Eqs. (\ref%
{ConsEqGB}) and (\ref{EnergyGB}) become
\begin{equation}
\dot{\rho }+2H(\rho +P)=0,  \label{ConsEqGBChoice}
\end{equation}
\begin{equation}
\epsilon =\frac{c\gamma }{g(\varepsilon)^{4}}\rho ^{\frac{\gamma -2}{\gamma }%
}.  \label{EnergyGBChoice}
\end{equation}

Now, substituting Eqs. (\ref{EoS}) and (\ref{FEQGBChoice}) in Eq. (\ref%
{ConsEqGBChoice}), we obtain
\begin{equation}
\dot{\rho }=\pm 2\gamma \rho \left[ \frac{1}{\alpha f(\rho )^{2}}\left( \pm
\sqrt{1+\frac{32}{3}\pi G\rho \alpha }-1\right) \right] ^{\frac{1}{2}},
\label{ConsVSf}
\end{equation}
where we will choose the plus sign in parenthesis to obtain a consistent
equation in Einstein gravity, i.e., in the limit $\alpha \longrightarrow 0$.
We expressed $f$ as a function of $\rho $ instead of $\epsilon$ according to
Eq. (\ref{EnergyGBChoice}). Now, one can show that finite-time singularities
(including big bang singularity) are absent if $f$ grows asymptotically as $%
\rho ^{1/4}$, or faster. For example, if $f\sim \rho ^{s}$, where $s\geq 1/4$%
. In this case, one can calculate the time for reaching a potential
singularity by integrating Eq. (\ref{ConsVSf}) (starting from some initial
finite density $\rho^{\ast}$ to an infinite one). This integration leads to
\begin{equation}
t=\pm \frac{\sqrt{\alpha }}{2\gamma }\int_{\rho ^{\ast }}^{\infty }\rho
^{s-1}\left( \sqrt{1+\frac{32}{3}\pi G\rho \alpha }-1\right) ^{-\frac{1}{2}%
}d\rho .  \label{IntChoice}
\end{equation}

After some calculations we obtain
\begin{equation}
t=\pm \left. \frac{\rho ^{s}\left\{ 2(16s^{2}-32s+15)\mathcal{H}_{1}+\varrho
\right\} }{\gamma (4s-1)(4s-3)(4s-5)(1+\mathcal{X})^{s}}\sqrt{\frac{\alpha
\mathcal{X}}{2}}\right\vert _{\rho ^{\ast }}^{\infty },  \label{TimeChoice}
\end{equation}%
where%
\begin{eqnarray*}
&&\varrho =-(4s-1)\mathcal{X}\left[ (4s-5)\mathcal{H}_{2}-(4s-3)\mathcal{XH}%
_{3}\right] ,
\end{eqnarray*}%
and $\mathcal{H}_{1}$, $\mathcal{H}_{2}, \mathcal{H}_{3}$ are the following
hypergeometric functions
\begin{eqnarray*}
&&\mathcal{H}_{1}={}_{2}F_{1}\left( \left[ -s,\frac{1}{2}-2s\right] ,\left[
\frac{3}{2}-2s\right] ,-\mathcal{X}\right) ,
\end{eqnarray*}%
\begin{eqnarray*}
&&\mathcal{H}_{2}={}_{2}F_{1}\left( \left[ -s,\frac{3}{2}-2s\right] ,\left[
\frac{5}{2}-2s\right] ,-\mathcal{X}\right) ,
\end{eqnarray*}%
\begin{eqnarray*}
&&\mathcal{H}_{3}={}_{2}F_{1}\left( \left[ -s+1,\frac{5}{2}-2s\right] ,\left[
\frac{7}{2}-2s\right] ,-\mathcal{X}\right) ,
\end{eqnarray*}
with
\begin{eqnarray*}
&&\mathcal{X}=\frac{6}{\sqrt{9+96\pi G\rho \alpha }-3}.
\end{eqnarray*}

Comparing various terms in Eq. (\ref{TimeChoice}), one can show that time is
infinite for $s\geq 1/4$. Thus, the time to reach the potential singularity
is infinite, so it is not a finite-time singularity, i.e., not physical. For
more clarification, we consider the term with $\rho $ as dominant term in
Eq. (\ref{IntChoice}), so we have
\begin{equation}
t=\pm \frac{\sqrt{\alpha }}{2\gamma }\int_{\rho ^{\ast }}^{\infty }\rho
^{s-1}\left( \frac{32}{3}\pi G\rho \alpha \right) ^{-\frac{1}{4}}d\rho
=\left. \pm \frac{2\sqrt{3\alpha }}{\gamma (4s-1)}\left( 96\pi G\alpha
\right) ^{-\frac{1}{4}}\rho ^{s-\frac{1}{4}}\right\vert _{\rho ^{\ast
}}^{\infty }=\infty ,\;s\geq \frac{1}{4},  \label{t-rhoGB}
\end{equation}%
and here we obtain the same result for $s$. We conclude that the rainbow
function $f(\varepsilon )$ plays an important role in possible resolution of
the big bang singularity, but it has to grow asymptotically as $\rho ^{1/4}$%
, or faster.

\begin{figure}[tbp]
$%
\begin{array}{cc}
\epsfxsize=7cm \epsffile{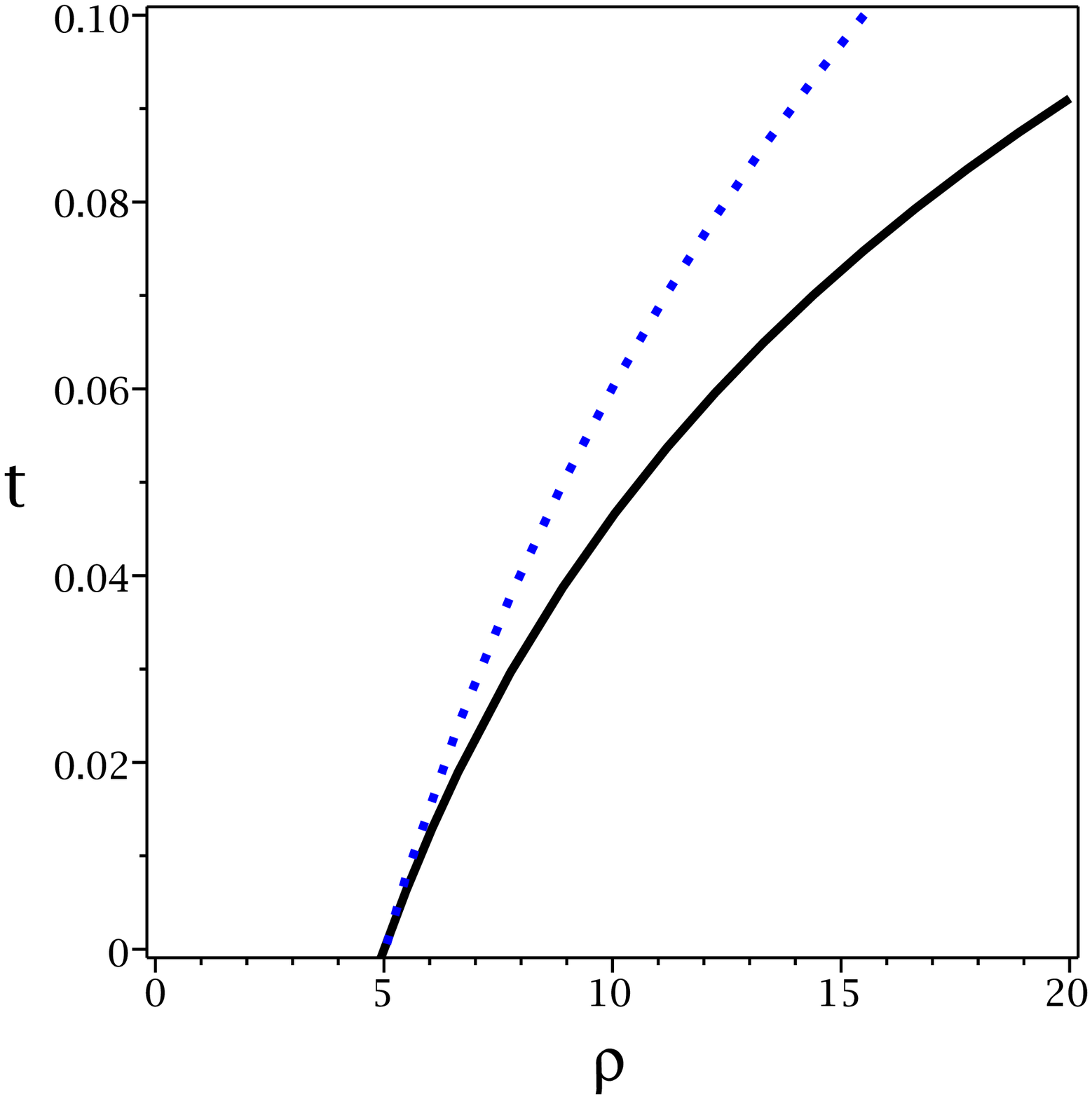} & \epsfxsize=7cm \epsffile{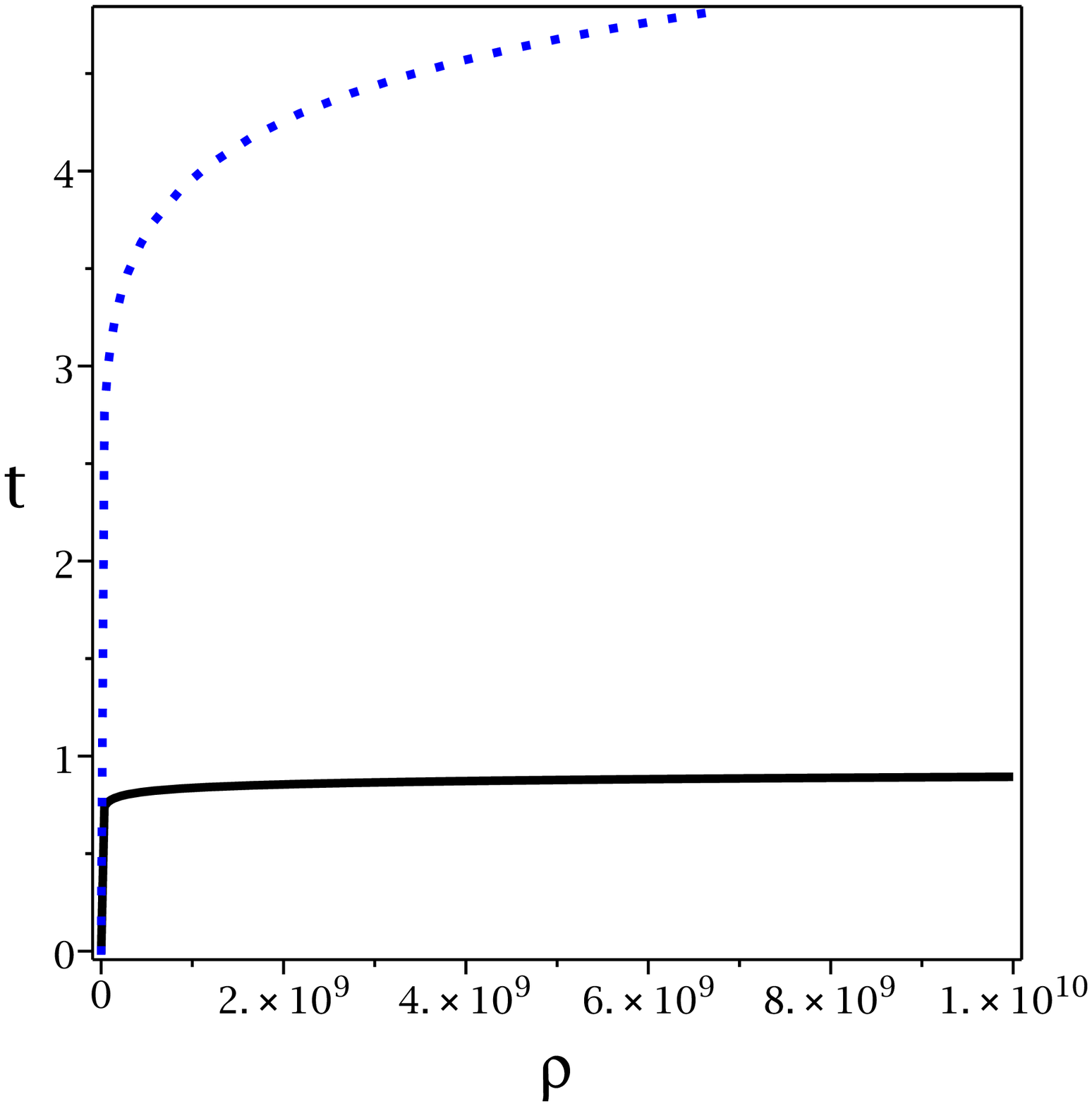}%
\end{array}
$%
\caption{\textbf{GB case:} Time versus density, for $G=2$, $\protect\gamma%
=4/3$, $\protect\alpha=0.1$, $\protect\rho ^{\ast }=5$, $s=1/5$ (continuous
line) and $s=1/3$ (doted line): \emph{"left and right figures indicate
various $\protect\rho$ ranges"}.}
\label{GBfig}
\end{figure}

Now, we can discuss Eq. (\ref{t-rhoGB}) and plot $t-\rho$ diagram for $s<1/4$
and $s>1/4$ in Fig. \ref{GBfig}. Considering this figure, one can find an
initial finite density at $t=0$ (present time), as expected. In addition,
for $s<1/4$, we obtain a finite value for time (to backward), when the
density of the universe goes to infinity (big bang singularity). However, in
the case of $s>1/4$, there is no finite (backward) time to obtain infinite
density and therefore, there is no big bang singularity at any finite time
in the past.

%%%%%%%%%%%%%%%%%%%%%%%%%%%%%%%%%%%%%%%%%%%%%%%%%%%%%%%%%%%%%%%%%%%%%%%%%%%%%%%%%%%%%%%

%%%%%%%%%%%%%%%%%%%%%%%%%%%%%%%%%%%%%%%%%%%%%%%%%%%%%%%%%%%%%%%%%%%%
%%%%%%%%%%%%%%%%%%%%%%%%%%%%%%%%%%%%%%%%%%%%%%%%%%%%%%%%%%%%%%%%%%%%

%%%%%%%%%%%%%%%%%%%%%%%%%%%%%%%%%%%%%%%%%%%%%%%%%%%%%%%%%%%%%%%%%%%%
\begin{figure}[tbp]
$%
\begin{array}{cc}
\epsfxsize=7cm \epsffile{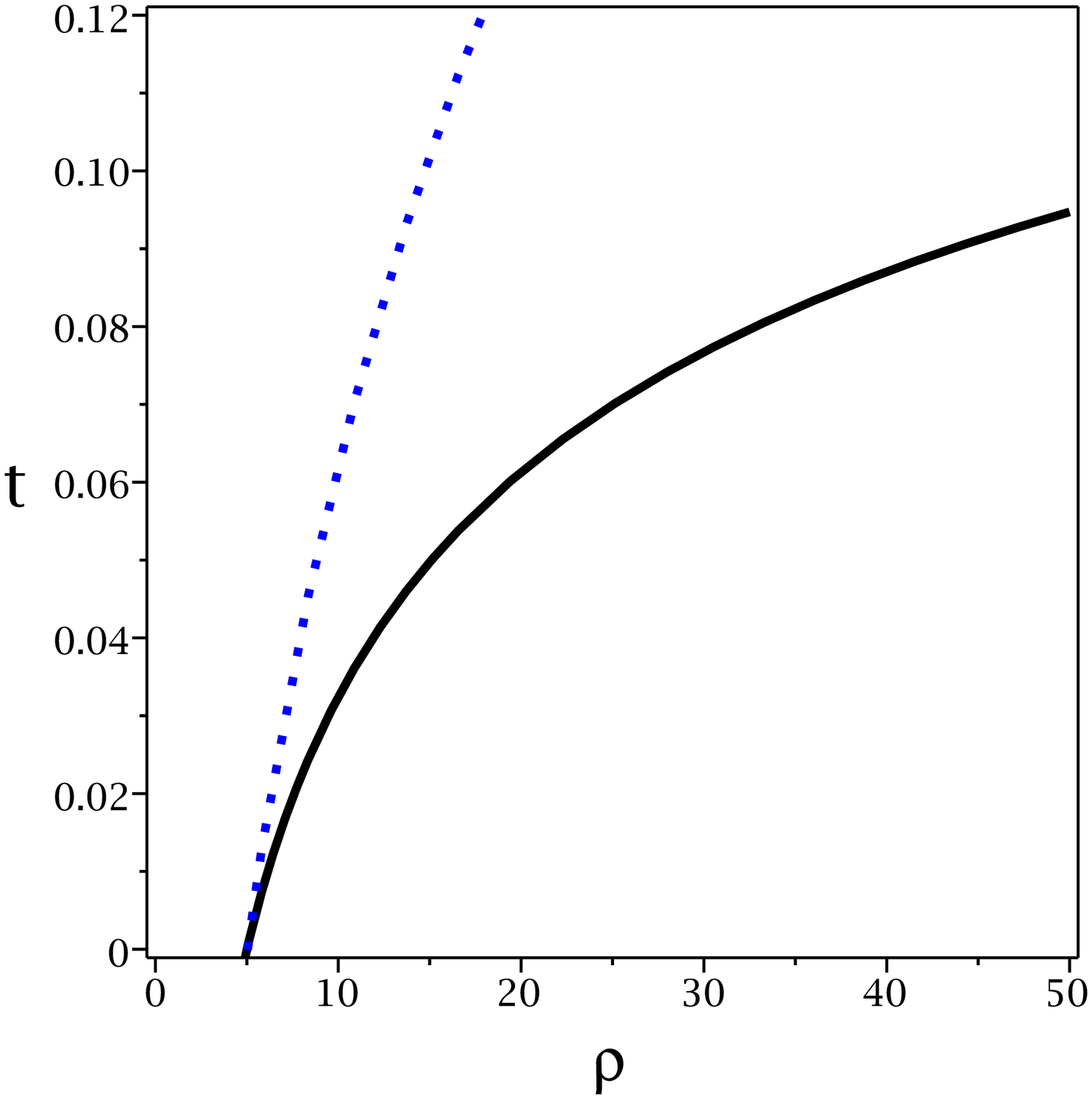} & \epsfxsize=7cm \epsffile{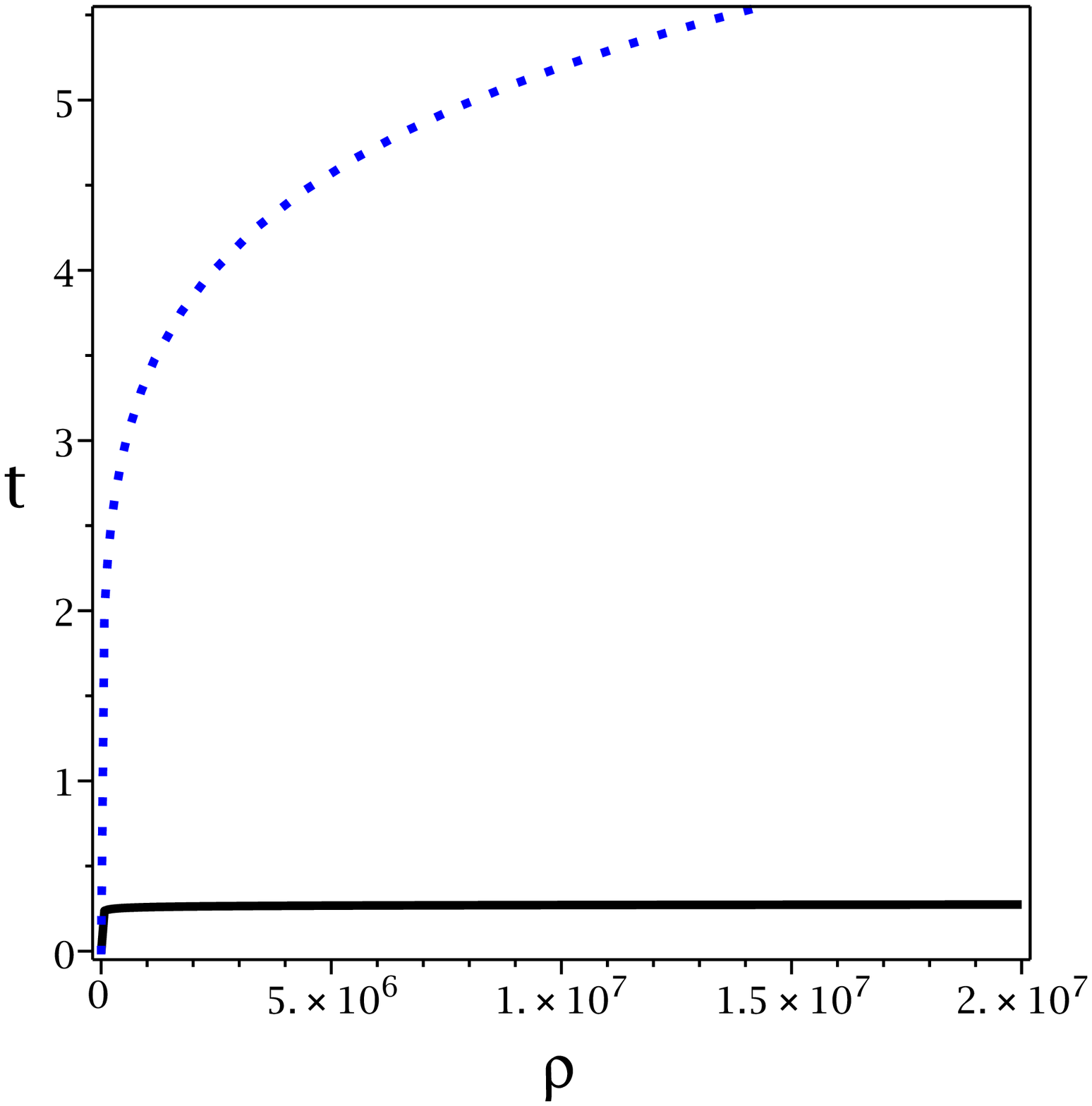}%
\end{array}
$%
\caption{\textbf{Einstein case:}Time versus density, for $G=2$, $\protect%
\gamma=4/3$, $\protect\rho ^{\ast }=5$, $s=1/3$ (continuous line) and $s=2/3$
(doted line): \emph{"left and right figures indicate various $\protect\rho$
ranges"}.}
\label{ENfig}
\end{figure}
%%%%%%%%%%%%%%%%%%%%%%%%%%%%%%%%%%%%%%%%%%%%%%%%%%%%%%%%%%%%%%%%%%%%
%%%%%%%%%%%%%%%%%%%%%%%%%%%%%%%%%%%%%%%%%%%%%%%%%%%%%%%%%%%%%%%%%%%%%%%%%%%%%%%%%%%%%%%

It is not necessary to perform similar analysis to investigate the rainbow
deformation of FRW cosmology in Einstein gravity. To do so, it is sufficient
to expand the function in Eq. (\ref{IntChoice}) for $\alpha \rightarrow 0$.
So, using rainbow deformation of the Einstein theory, we obtain
\begin{equation}
t=\pm \frac{1}{8\gamma }\sqrt{\frac{3}{\pi G}}\int_{\rho ^{\ast }}^{\infty
}\rho ^{s-\frac{3}{2}}d\rho =\left. \pm \frac{1}{4\gamma (2s-1)}\sqrt{\frac{3%
}{\pi G}}\rho ^{s-\frac{1}{2}}\right\vert _{\rho ^{\ast }}^{\infty }=\infty
,~\;s\geq \frac{1}{2},  \label{t-rhoEN}
\end{equation}%
where it shows that the value of $f$ for having a nonsingular universe in
Einstein gravity has to grow asymptotically as $\rho ^{1/2}$, or faster.

Here, we plot $t-\rho$ diagram in Einstein case (Eq. (\ref{t-rhoEN})) for
both $s<1/2$ and $s>1/2$ in Fig. \ref{ENfig}. Like GB case, Fig. \ref{ENfig}
shows that there is an initial finite density at $t=0$ (present time). There
is no infinite density (big bang singularity) in a finite (backward) time
for $s>1/2$. However, for $s<1/2$, there is a finite (backward) time at
which an infinite density (big bang singularity) exists.

Here, we have investigated possibility of obtaining a nonsingular rainbow
universe in the Einstein and GB gravities. In the coming section, we are
going to use the MDR (Eq. (\ref{MDR1})) to analyze such a nonsingular
FRW-like cosmology.

\section{NONSINGULAR RAINBOW UNIVERSES \label{NONSINGULAR}}

Using Eqs. (\ref{MDR}), (\ref{EoS}), and (\ref{ConsEqGB}), one can show that
average energy (\ref{EnergyGB}) can be expressed as
\begin{equation}
\epsilon =c\gamma \rho ^{\frac{\gamma -1}{\gamma }}.  \label{EnergyMDR}
\end{equation}

Now using Eqs. (\ref{MDR}) and (\ref{EnergyMDR}), the function $%
f(\varepsilon )$ will be
\begin{equation}
f(\epsilon )=\frac{\exp \left( \gamma \mathcal{G}^{\frac{\gamma -1}{\gamma }%
}\right) -1}{\gamma \mathcal{G}^{\frac{\gamma -1}{\gamma }}},
\end{equation}%
where $\mathcal{G}=\rho /\rho _{P}$, and $E_{P}=c\rho _{P}^{\frac{\gamma -1}{%
\gamma }}$ is the Planck energy versus density $\rho _{P}$. Using the above
equation and the MDR relation, one can show that the modified Friedmann
equation (\ref{FEQGB1}) will be given by
\begin{equation}
H=\pm \frac{\gamma \mathcal{G}^{\frac{\gamma -1}{\gamma }}\left[ -9\pm 3%
\sqrt{9+96\pi \alpha \mathcal{G}\rho _{P}^{\frac{2-\gamma }{\gamma }}}\right]
^{\frac{1}{2}}}{6\sqrt{\alpha }\left[ \exp \left( \gamma \mathcal{G}^{\frac{%
\gamma -1}{\gamma }}\right) -1\right] }.  \label{FeqMDR}
\end{equation}%
We will choose plus sign in parenthesis to obtain a consistent equation in
Einstein gravity, i.e., in the limit $\alpha \longrightarrow 0$. We can
investigate a possible singular solution of the big bang singularity using
the discussion of section \ref{POSSIBILITY}. Substituting Eq. (\ref{EoS})
and the modified Friedmann equation (\ref{FeqMDR}) in Eq. (\ref{ConsEqGB})
and using Eq. (\ref{MDR}), we can obtain the following equation
\begin{equation}
\dot{\mathcal{G}}=\pm \frac{2\gamma ^{2}\mathcal{G}^{\frac{2\gamma -1}{%
\gamma }}\left[ -9+3\sqrt{9+96\pi \alpha \mathcal{G}\rho _{P}^{\frac{%
2-\gamma }{\gamma }}}\right] ^{\frac{1}{2}}}{3\sqrt{\alpha }\left[ \exp
\left( \gamma \mathcal{G}^{\frac{\gamma -1}{\gamma }}\right) -1\right] },
\label{ConsVSfMDR}
\end{equation}%
where $\dot{\mathcal{G}}=\dot{\rho}/\rho _{P}$.

Now, we want to show that the time is infinite when we go from an initial
finite density $\mathcal{G}^{\ast }$ to an infinite one in special case $%
\gamma =4/3$, (i.e., radiation). This can be done by integrating Eq. (\ref%
{ConsVSfMDR})
\begin{equation}
t=\pm \frac{27\sqrt{\alpha }}{32}\int_{\mathcal{G}^{\ast }}^{\infty }\frac{%
\exp \left( \frac{4}{3}\mathcal{G}^{\frac{1}{4}}\right) -1}{\mathcal{G}^{%
\frac{5}{4}}\left[ -9+3\sqrt{9+96\pi \alpha \mathcal{G}\rho _{P}^{\frac{1}{2}%
}}\right] ^{\frac{1}{2}}}d\mathcal{G},  \label{TimeMDR}
\end{equation}%
in which it is too hard to compute this integration analytically; however
one can use numerical calculation to show that it does not converge on $%
\left[ \mathcal{G}^{\ast },\infty \right) $, so the time to reach the
infinite density is infinite. For more clarification, we consider the term
with $\mathcal{G}$ as dominant term in denominator, and so we obtain
\begin{eqnarray}
&&t=\pm \frac{27}{32}\left[ \frac{3}{\alpha }\sqrt{96\pi \alpha \rho _{P}^{%
\frac{1}{2}}}\right] ^{-\frac{1}{2}}\int_{\mathcal{G}^{\ast }}^{\infty }%
\frac{\exp \left( \frac{4}{3}\mathcal{G}^{\frac{1}{4}}\right) -1}{\mathcal{G}%
^{\frac{3}{2}}}d\mathcal{G}  \nonumber \\
&&  \nonumber \\
&=&\left. \pm \frac{27}{32}\left[ \frac{3}{\alpha }\sqrt{96\pi \alpha \rho
_{P}^{\frac{1}{2}}}\right] ^{-\frac{1}{2}}\left( \frac{2}{\mathcal{G}^{\frac{%
1}{2}}}-\frac{32\mathcal{E}\left( 1,-\frac{4}{3}\mathcal{G}^{\frac{1}{4}%
}\right) }{9}-\frac{2\mathcal{G}^{\frac{-1}{2}}\left( 3+4\mathcal{G}^{\frac{1%
}{4}}\right) }{3\exp \left( \frac{-4\mathcal{G}^{\frac{1}{4}}}{3}\right) }%
\right) \right\vert _{\mathcal{G}^{\ast }}^{\infty }=\infty ,  \label{Time2}
\end{eqnarray}%
where $\mathcal{E}$ is the exponential integration. This shows that the time
to reach this infinite density is infinite. Thus, there is no finite-time
singularities, and this result confirms the consequence of Eq. (\ref{TimeMDR}%
).

In order to investigate the rainbow deformation of the Einstein gravity, one
can follow the same procedure using Eqs. (\ref{MDR}), (\ref{FEQE1}), (\ref%
{ConsEqE}) and (\ref{EnergyMDR}) or just expand the function in integration (%
\ref{TimeMDR}) for $\alpha \rightarrow 0$. Here, we use the second way, and
we obtain
\begin{equation}
t=\pm \frac{9}{128}\sqrt{\frac{3}{\pi \rho _{P}^{\frac{1}{2}}}}\int_{%
\mathcal{G}^{\ast }}^{\infty }\frac{\exp \left( \frac{4}{3}\mathcal{G}^{%
\frac{1}{4}}\right) -1}{\mathcal{G}^{\frac{7}{4}}}d\mathcal{G}=\left. \pm
\frac{9}{128}\sqrt{\frac{3}{\pi \rho _{P}^{\frac{1}{2}}}}\left( \frac{4}{3%
\mathcal{G}^{\frac{3}{4}}}+\zeta \right) \right\vert _{\mathcal{G}^{\ast
}}^{\infty }=\infty ,  \label{TimeMDRE}
\end{equation}%
where%
\[
\zeta =-\frac{128\mathcal{E}\left( 1,-\frac{4}{3}\mathcal{G}^{\frac{1}{4}%
}\right) }{81}-\frac{4\left( 9+8\mathcal{G}^{\frac{1}{2}}+6\mathcal{G}^{%
\frac{1}{4}}\right) }{27\mathcal{G}^{\frac{3}{4}}\exp \left( \frac{-4}{3}%
\mathcal{G}^{\frac{1}{4}}\right) }.
\]%
This result shows that there is no finite-time singularities.

\begin{figure}[tbp]
$%
\begin{array}{c}
\epsfxsize=7cm \epsffile{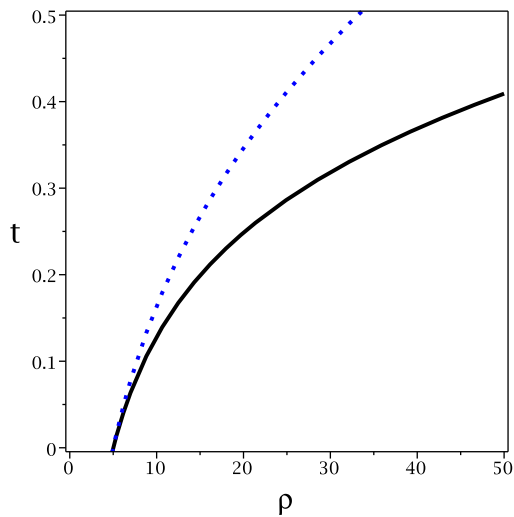}%
\end{array}
$%
\caption{Time versus density, for $\mathcal{G}^{\ast }=5$ and $\protect\rho%
_{P}=0.2$. \newline
Einstein gravity's rainbow ($\protect\alpha=0$: continuous line) and
Gauss-Bonnet gravity's rainbow ($\protect\alpha=0.1$: doted line).}
\label{ENGB}
\end{figure}

We also plot $t-\rho$ diagram for both Einstein and GB gravities (Eqs. (\ref%
{Time2}) and (\ref{TimeMDRE})) in Fig. \ref{ENGB}. This figure shows that
for both Einstein and GB gravities, there is an initial finite density at $%
t=0$, however, (backward) time goes to infinity as density goes to infinity
and therefore there is no big bang singularity.

Finally, it is interesting to investigate the behavior of density of states
at the Planck scale to analyze its divergences \cite{Ling,LingW}. Using the
MDR, the density of states can be written as
\begin{equation}
a(E)dE\simeq p^{3}dp\simeq f(\varepsilon)^{4}\left( 1+E\frac{%
f(\varepsilon)^{\prime }}{f(\varepsilon)} \right) E^{3}dE.
\end{equation}
Here by substituting the MDR and using the fact that energy cannot be larger
than the Planck energy, the density of states has a finite value $e(e-1)^{3}$
with regular behavior without any divergences.

%%%%%%%%%%%%%%%%%%%%%%%%%%%%%%%%%%%%%%%%%%%%%%%%%%%%%%%%%%%%%%%%%%%%
%%%%%%%%%%%%%%%%%%%%%%%%%%%%%%%%%%%%%%%%%%%%%%%%%%%%%%%%%%%%%%%%%%%%

\section{Conclusions \label{Conclusions}}

In this work, we have investigated the effect of gravity's rainbow in
Einstein and GB gravities for the early universe. We have analyzed the
rainbow deformation of the $5-$ dimensional FRW solution in both Einstein
and GB gravity. We have observed that although GB term contributes to the
field equations, it does not change the conservation equation and average
energy. We have also demonstrated that the rainbow functions modify both
conservation equation and average energy. Then, we have discussed the
general conditions for having the nonsingular FRW cosmologies using the
rainbow deformation of both Einstein and GB gravities.

We have used the rainbow functions defined by Amelino-Camelia, et. al., \cite%
{Amelinod,AmelinoEMNS} to investigate the effect of rainbow deformation of
FRW-like cosmology. We have demonstrated that it is possible to obtain
nonsingular cosmological solutions by using a rainbow deformation of
Einstein and GB gravities. The Friedmann equations were modified using the
gravity's rainbow by suitable rainbow functions. We also identified the
rainbow functions with the MDR introduced by Amelino-Camelia, et al. \cite%
{Amelinod,AmelinoEMNS}, and studied the rainbow modified Friedmann equations
of a perfect fluid. We have found nonsingular solutions for a wide range of
values for the equation of state parameter $\gamma >4/3$ in both Einstein
and GB gravities. We also found that GB gravity has considerable effect on
the constraint for having nonsingular universes. Using the analysis done in
\cite{AwadAM}, we found that the universe takes infinite time to reach $\rho
\rightarrow \infty $ from a finite value of $\rho $. We have also found that
the density of states do not diverge at the Planck scale. So for both cases,
we found a possible resolution of the big bang singularity. Hence, it seems
that the removal of singularities by rainbow deformation is not a model
dependent effect. It would be interesting to perform this analysis in other
models of Lovelock gravity.

\acknowledgments

We would like to thank the anonymous referee for useful suggestions and
enlightening comments. We also acknowledge S. Panahiyan and Z. Armanfard for
reading the manuscript. SHH, MM and BE wish to thank Shiraz University
Research Council. This work has been supported financially by the Research
Institute for Astronomy and Astrophysics of Maragha, Iran.

%\clearpage

%\clearpage

%\begin{figure}
%\epsscale{1.1}
%\plotone{fig1.eps}
%\caption{Travel-time differences, $\delta\tau_{\rm{\rm{NS}}}$, against latitude for four selected travel distances ($\Delta$). Each measurement shown here is an average of
%individual measurements, with a spacing $0.72$ deg. in $\Delta$ and one for each of 47 months. The error bars represent
%standard errors estimated from these individual measurements. The center-to-limb systematics in travel times estimated through $\delta\tau_{\rm{WE}}$ have been subtracted from $\delta\tau_{\rm{NS}}$.}
%\label{fig1}
%\end{figure}

\end{document}